\documentclass[aps,pra,twocolumn,nofootinbib,preprintnumbers,english]{revtex4-1}
\usepackage{amsthm,amsmath,amsfonts,dsfont,upgreek}    
\usepackage{amssymb,epsfig,setspace}
\usepackage{graphicx}
\usepackage{babel}
\usepackage{verbatim}

\begin{document}

\hyphenation{ano-ther ge-ne-ra-te dif-fe-rent know-le-d-ge po-ly-no-mi-al}
\hyphenation{me-di-um  or-tho-go-nal as-su-ming pri-mi-ti-ve pe-ri-o-di-ci-ty}
\hyphenation{mul-ti-p-le-sca-t-te-ri-ng i-te-ra-ti-ng e-q-ua-ti-on}
\hyphenation{wa-ves di-men-si-o-nal ge-ne-ral the-o-ry sca-t-te-ri-ng}
\hyphenation{di-f-fe-r-ent tra-je-c-to-ries e-le-c-tro-ma-g-ne-tic pho-to-nic}
\hyphenation{Ray-le-i-gh di-n-ger Kra-jew-ska Wal-c-zak Ham-bur-ger Ad-di-ti-o-nal-ly}
\hyphenation{Kon-ver-genz-the-o-rie ori-gi-nal in-vi-si-b-le cha-rac-te-ri-zed}
\hyphenation{Ne-ver-the-less sa-tu-ra-te Ene-r-gy sa-ti-s-fy}

\title{On solvability and integrability of the Rabi model}

\author{Alexander Moroz}

\affiliation{Wave-scattering.com} 
 
\begin{abstract}
Quasi-exactly solvable Rabi model is investigated within the
framework of the Bargmann Hilbert space of analytic functions 
${\cal B}$. On applying the theory of orthogonal polynomials, 
the eigenvalue equation and eigenfunctions are shown to be determined 
in terms of three systems of monic orthogonal polynomials. 
The formal Schweber quantization criterion for an energy variable 
$x$, originally expressed in terms of infinite continued fractions, 
can be recast in terms of a meromorphic function 
$F(z) = a_0 + \sum_{k=1}^\infty {\cal M}_k/(z-\xi_k)$ in the complex 
plane $\mathbb{C}$ with {\em real simple} poles $\xi_k$
and {\em positive} residues ${\cal M}_k$.
The zeros of $F(x)$ on the real axis determine the spectrum of the Rabi model.
One obtains at once that, on the real axis, 
(i) $F(x)$ monotonically decreases from $+\infty$ to $-\infty$ 
between any two of its subsequent poles $\xi_k$ and $\xi_{k+1}$,
(ii) there is exactly one zero of $F(x)$ for $x\in (\xi_k,\xi_{k+1})$, 
and (iii) the spectrum corresponding to the zeros of $F(x)$ 
does not have any accumulation point. Additionally, one can 
provide much simpler proof of that the spectrum in 
each parity eigenspace  ${\cal B}_\pm$ is necessarily {\em nondegenerate}.
Thereby the calculation of spectra is greatly facilitated. 
Our results allow us to critically examine recent claims regarding 
solvability and integrability of the Rabi model. 
\end{abstract}


\maketitle 

%

\section{Introduction}
\label{sc:intr}
The Rabi model \cite{Rb} describes the simplest interaction between a 
cavity mode with a bare frequency $\omega$ and a two-level system 
with a bare resonance frequency $\omega_0$. 
The model is characterized by the Hamiltonian \cite{Rb,Schw,Br,YZL} 
\begin{equation}
\hat{H}_R =
\hbar \omega \mathds{1} \hat{a}^\dagger \hat{a}  
 +  \hbar g\sigma_1 (\hat{a}^\dagger + \hat{a}) + \mu \sigma_3,
\label{rabih}
\end{equation}
where $\mathds{1}$ is the unit matrix, 
$\hat{a}$ and $\hat{a}^\dagger$ are 
the conventional boson annihilation and creation operators 
satisfying commutation relation 
$[\hat{a},\hat{a}^{\dagger}] = 1$,
$g$ is a coupling constant, and $\mu=\hbar \omega_0/2$. 
In what follows we assume the standard representation
of the Pauli matrices $\sigma_j$ and set the 
reduced Planck constant $\hbar=1$.
For dimensionless coupling strength $\kappa=g/\omega \lesssim 10^{-2}$,
the physics of the Rabi model is well captured by the
analytically solvable approximate 
Jaynes and Cummings (JC) model \cite{JC,SK}. 
The latter is obtained from the former upon applying 
the rotating wave approximation (RWA), whereby the 
coupling term 
$\sigma_1 (\hat{a}^\dagger + \hat{a})$ in Eq. (\ref{rabih}) is replaced by 
$(\sigma_+ \hat{a} + \sigma_-\hat{a}^\dagger)$, 
where $\sigma_\pm \equiv (\sigma_1 \pm i \sigma_2)/2$.
Nowdays, solid-state semiconductor \cite{KGK} 
and superconductor systems \cite{BGA,FLM,NDH} have allowed
the advent of the {\em ultrastrong} coupling  regime,
where the dimensionless coupling strength $\kappa\gtrsim 0.1$ \cite{CC}. 
In this regime, the validity of the RWA breaks down and
the relevant physics can only be 
described by the full Rabi model \cite{Rb}. 
With new experiments rapidly approaching the limit of
the {\em deep strong} coupling regime 
characterized in that $\kappa\gtrsim 1$ \cite{CRL}, i.e.,
an order of magnitude stronger coupling,
the relevance of the Rabi model \cite{Rb} becomes even more
prominent. There is every reason to believe that 
ultrastrong and deep strong coupling  systems could 
open up a rich vein of research on truly 
quantum effects with implications for quantum 
information science and fundamental quantum optics \cite{KGK}.

The Rabi model applies to a great variety of physical systems, 
including cavity and circuit quantum electrodynamics, quantum
dots, polaronic physics and trapped ions. In spite 
of recent claims \cite{Br,Sln}, the model is {\em not} 
exactly solvable. Rather it is a typical
example of {\em quasi-exactly solvable} (QES) models 
in quantum mechanics \cite{TU,Trb,BD,KUW,KKT}. 
The QES models are distinguished by the fact 
that a {\em finite} number of their eigenvalues and 
corresponding eigenfunctions can be determined algebraically
\cite{TU,Trb,BD,KUW}. 
That is also the case of the Rabi model \cite{KKT}.
Certain energy levels of the Rabi model, known 
as Juddian exact isolated
solutions \cite{Jd}, can be analytically
computed \cite{Jd,Ks,KL}, whereas the remaining part of the
spectrum not \cite{Ks,KL}. 
Depending on model parameters, the spectrum can only be 
approximated (sometime rather accurately - 
cf. Eqs. (18), (20) and Fig. 3 of Ref. \cite{FKU}; 
Eq. (20) and Figs. 1,2 of Ref. \cite{Ir}).
Therefore, any kind of exact results involving
the Rabi model continues to be of great theoretical and experimental 
value.

In our earlier work \cite{AMep} we studied 
the Rabi model as a member of a more general class ${\cal R}$ of 
quantum models. 
In the Hilbert space ${\cal B}=L^2(\mathbb{R})\otimes\mathbb{C}^2$, where
$L^2(\mathbb{R})$ is represented by the Bargmann space of 
entire functions $\mathfrak{b}$, and $\mathbb{C}^2$ 
stands for a spin space \cite{Schw,Brg}, 
the models of the class ${\cal R}$ were characterized in that 
the eigenvalue equation    
\begin{equation}
\hat{H}\Phi=E\Phi,
\label{eme}
\end{equation}
where $\hat{H}$ denotes a corresponding Hamiltonian, reduces 
to a {\em three-term difference equation} 
\begin{equation}
\phi_{n+1} + a_n \phi_n + b_n \phi_{n-1}=0 \hspace*{1.2cm}  (n\ge 0).
\label{3trg}
\end{equation}
Here $\{\phi_n\}_{n=0}^\infty$ are the sought 
expansion coefficients of an entire function 
\begin{equation}
\upvarphi(z)=\sum_{n=0}^\infty \phi_n z^n
\label{pss}
\end{equation}
in $\mathfrak{b}$ that generates
a physical state $\Phi(z)$ (in general vector 
in a spin space - see below).
Models of the class ${\cal R}$ were then 
characterized in that the recurrence coefficients
have an asymptotic powerlike dependence \cite{AMep}
\begin{equation}
a_n\sim a n^{\varsigma},~~~~~~~~~~~ 
          b_n\sim b n^{\upsilon}\hspace*{1.2cm} (n\rightarrow\infty),
\label{rcd}
\end{equation}
where $a$ and $b$ are proportionality constants
and the exponents satisfy 
$2\varsigma>\upsilon$ and $\tau=\varsigma-\upsilon\geq 1/2$ \cite{AMep}.
In virtue of the Perron and Kreuser generalizations 
(Theorems 2.2 and 2.3(a) in Ref. \cite{Gt}) of the Poincar\'{e} theorem 
(Theorem 2.1 in Ref. \cite{Gt}), the recurrence 
equation (\ref{3trg}) (considered for $n\ge 1$) possesses 
{\em two} linearly independent solutions:
\begin{itemize}

\item (i) a {\em dominant} solution $\{d_j\}_{j=0}^\infty$ and

\item  (ii) a {\em minimal} solution $\{m_j\}_{j=0}^\infty$.
\end{itemize}

The respective solutions differ in the 
behavior $\phi_{n+1}/\phi_n$ in the limit $n\rightarrow\infty$.
The {\em minimal} solution guaranteed by 
the Perron-Kreuser theorem
(Theorem 2.3 in Ref. \cite{Gt}) for models 
of the class ${\cal R}$ satisfies 
\begin{equation}
\frac{m_{n+1}}{m_n}\sim -\frac{b}{a}\frac{1}{n^\tau} \rightarrow 0
\hspace*{1.2cm} (n\rightarrow \infty)
\label{mins}
\end{equation}
in virtue of (\ref{rcd}) and $\tau \geq 1/2>0$.
On substituting the minimal solution 
for the $\phi_n$'s in Eq. (\ref{pss}), 
$\upvarphi(z)$ automatically becomes an {\em entire} function
belonging to $\mathfrak{b}$ \cite{AMep}.
In what follows, only the minimal solutions will be considered,
and $\upvarphi(z)$ in Eq. (\ref{pss}) will stand for the entire function 
generated by the minimal solution [of the $n\ge 1$ part 
of Eq. (\ref{3trg})].

The spectrum of any quantum model of ${\cal R}$ 
can be obtained as zeros of 
the transcendental function of a dimensionless energy 
parameter $x$ \cite{AMep,AMcm}, 
\begin{equation}
F(x)\equiv a_0 + \sum_{k=1}^\infty \rho_1\rho_2\ldots\rho_k,
\label{fdf}
\end{equation}
where $F(x)$ is defined solely in terms of the coefficients of 
the three-term recurrence \cite{AMep,AMcm,AMr}:
\begin{eqnarray}
\rho_1 &=& -\frac{b_1}{a_1},
\hspace*{0.3cm}
\rho_l=u_l-1,
\hspace*{0.3cm}
u_1=1,
\nonumber\\
u_l &=& \frac{1}{1-u_{l-1}b_l/(a_la_{l-1})},
\hspace*{0.3cm}
l\geq 2.
\label{eth}
\end{eqnarray}
The function $F(x)$ 
yields both analytic and efficient numerical representation of the
formal Schweber quantization criterion 
expressed in terms of infinite  continued
fractions (cf. Eq. (A.16) of Ref. \cite{Schw}),
\begin{equation}
0 = F(x)\equiv a_0 + \frac{-b_{1}}{a_{1}-} \frac{b_{2}}{a_{2}-}
\frac{b_{3}}{a_{3}-}\cdots
\label{fdfs}
\end{equation}
The condition $F(x)=0$ is equivalent to 
that the logarithmic derivative of 
$\upvarphi(z)$ satisfies the {\em boundary condition} 
\begin{equation}
r_{0}\equiv \frac{\upvarphi'(0)}{\upvarphi(0)}=-a_0,
\label{bcc}
\end{equation}
where $r_{0}$ stands for the infinite continued fraction
on the right-hand side in Eq. (\ref{fdfs}) \cite{AMep}.
An important insight missed in Refs. \cite{Schw,AMep}
is that the spectrum of any quantum model of ${\cal R}$ 
is necessarily {\em nondegenerate}, unless
the conditions that guarantee uniqueness
of the minimal solution the recurrence 
(\ref{3trg}) cannot be satisfied
(cf. Sec. \ref{sc:dgnr}).

The Rabi Hamiltonian $\hat{H}_R$ is known
to possess a discrete $\mathbb{Z}_2$-symmetry corresponding 
to the constant of motion, or parity, 
$\hat{\Pi}=\exp(i\pi \hat{J})$  \cite{Br,CRL,Ks}, where
\begin{equation}
\hat{J}=\mathds{1} \hat{a}^\dagger \hat{a}
         +\frac{1}{2}\,(\mathds{1}+\sigma_3)
\label{jop}
\end{equation}
is the familiar operator known to generate a continuous $U(1)$
symmetry of the JC model \cite{Br,JC}. 
In order to employ the Fulton and Gouterman 
reduction \cite{FG} in the 
{\em positive} and {\em negative parity spaces}, 
wherein one component 
of $\Phi$ is generated from the other by means of 
a suitable cyclic operator $\hat{\gamma}$, $\hat{\gamma}^2=1$,
it is expedient to work in a
unitary equivalent {\em single-mode spin-boson picture}
\begin{equation}
\hat{H}_{sb} =
\omega \mathds{1} \hat{a}^\dagger \hat{a} + \mu \sigma_1  
         + g \sigma_3 (\hat{a}^\dagger + \hat{a}).
\nonumber
\end{equation}
The transformation is accomplished 
by means of the unitary operator 
$U=(\sigma_1 + \sigma_3)/\sqrt{2} = U^{-1}$.
Hamiltonian $\hat{H}_{sb}$  is then 
of the Fulton and Gouterman type (see Sec. IV of Ref. \cite{FG})
\begin{equation}
\hat{H}_{FG}= A\mathds{1} + B\sigma_1 + C\sigma_3,
\nonumber
\end{equation}
with
\begin{equation}
A=\omega  \hat{a}^\dagger \hat{a},~~~~~
B=\mu,~~~~~
C=g (\hat{a}^\dagger + \hat{a}).
\nonumber
\end{equation}
The Fulton and Gouterman symmetry operation 
is realized by $\hat{\gamma}=e^{i\pi \hat{a}^\dagger \hat{a}}$, 
which transforms a given operator $\hat{O}$ according to
\begin{equation}
\hat{O}\rightarrow 
e^{i\pi \hat{a}^\dagger \hat{a}} 
       \hat{O} e^{-i\pi \hat{a}^\dagger \hat{a}}.
\nonumber
\end{equation}
The latter induces {\em reflections} 
of the annihilation and creation operators:
$\hat{a}\rightarrow-\hat{a}$, 
$\hat{a}^\dagger\rightarrow-\hat{a}^\dagger$, 
and leaves the boson number operator 
$\hat{a}^\dagger \hat{a}$ invariant \cite{Br,FG}.
Because $[\hat{\gamma},A]=[\hat{\gamma},B]=\{ \hat{\gamma},C\}=0$, 
$\hat{\Pi}_{FG}=\sigma_1 \hat{\gamma}$ is the 
symmetry of $\hat{H}_{sb}$ \cite{Br,FG}.
One can verify that, 
with $\sigma_3$ replaced by $\sigma_1$ in Eq. (\ref{jop}),
\begin{equation}
\hat{\Pi}= e^{i\pi \hat{J}} = - \sigma_1 e^{i\pi \hat{a}^\dagger \hat{a}} 
            = -\sigma_1 \hat{\gamma} =-\hat{\Pi}_{FG}. 
\nonumber
\end{equation}
Such as to any cyclic $\mathbb{Z}_2$ operator, one can associate 
to $\hat{\Pi}_{FG}$ a pair of {\em projection operators}
\begin{equation}
P^\pm = \frac{1}{2}\, ( \mathds{1} \pm \hat{\Pi}_{FG}),
              ~~~~~\left(P^\pm\right)^2 = P^\pm.
\nonumber
\end{equation}
The respective projection operators $P^\pm$ 
project out eigenstates of $\hat{\Pi}_{FG}$:
an arbitrary state $\Psi\in {\cal B}$ is projected into 
corresponding parity eigenstates $\Phi^\pm$ 
with positive and negative parity. 
In the conventional {\em off-diagonal} Pauli representation of 
$\sigma_1$ one has \cite{FG}:
\begin{equation}
P^+ \Psi = \frac{1}{2}\,
\left(
\begin{array}{cc}
1 & \hat{\gamma} \\
\hat{\gamma} & 1
\end{array}\right)\left(
\begin{array}{c}
\uppsi_1 \\
\uppsi_2
\end{array}\right) =
\frac{1}{2}\,
\left(
\begin{array}{c}
\uppsi_1 + \hat{\gamma} \uppsi_2
\\
\hat{\gamma}(\uppsi_1 + \hat{\gamma} \uppsi_2)
\end{array}\right). 
\label{fgb}
\end{equation}
The right-hand side of Eq. (\ref{fgb}) shows that one
component of the positive parity eigenstate 
can be generated from the other by 
means of the symmetry operator $\hat{\gamma}$ \cite{FG}. 
A similar argument holds for $P^-$, wherein $-\hat{\gamma}$ is 
substituted for $\hat{\gamma}$ in Eq. (\ref{fgb}).
Therefore, the corresponding 
parity eigenstates $\Phi^+$ and $\Phi^-$ of the eigenvalue
equation (\ref{eme}) 
contain one independent component each,
\begin{equation}
\Phi^+(z)  = \left(
\begin{array}{c}
\upvarphi^+
\\
\hat{\gamma}\upvarphi^+
\end{array}\right), \hspace*{0.8cm}
\Phi^-(z)  = \left(
\begin{array}{c}
\upvarphi^-
\\
- \hat{\gamma}\upvarphi^-
\end{array}\right).
\label{fgr}
\end{equation} 
For the sake of comparison with Ref. \cite{Br},
the superscript $\pm$ denotes the {\em positive} and {\em negative} 
parity states of $\hat{\Pi}_{FG}$ and not of 
the conventional parity operator $\hat{\Pi}$.

The respective parity eigenstates $\Phi^+(z)$ and $\Phi^-(z)$ 
satisfy the following eigenvalue equations for the 
independent (e.g. upper) component 
(cf. Eqs. (4.12-13) of Ref. \cite{FG})
\begin{eqnarray}
H^+\upvarphi^+ &=& [A + B\hat{\gamma} + C ]\upvarphi^+= E^+\upvarphi^+,
\nonumber
\\
H^-\upvarphi^- &=& [A - B\hat{\gamma} + C ]\upvarphi^-= E^-\upvarphi^-.
\nonumber
\end{eqnarray}
Here we have written $E^\pm$ since, in general, the spectra 
of $H^+$ and $H^-$ do not coincide.
In the Bargmann space of entire functions 
$\mathfrak{b}$, the action of $\hat{\gamma}$ becomes \cite{AMep}
\begin{equation}
\hat{\gamma}\upvarphi^\pm (z)=\upvarphi^\pm (-z) 
           =\sum_{n=0}^\infty (-1)^n \phi_n^\pm z^n.
\nonumber
\end{equation}
Therefore, the Rabi model can be characterized 
by a pair of the three-term recurrences (Eq. (37) of Ref. \cite{AMep})
\begin{eqnarray}
 \lefteqn{
 \phi_{n+1}^\pm +\frac{1}{\kappa (n+1)}\, 
              [n  - \upepsilon  \pm(-1)^n\Delta]\phi_{n}^\pm 
}\hspace*{3cm}
\nonumber\\
      &&    + \frac{1}{n+1}\, \phi_{n-1}^\pm = 0,
\label{rbmb2}
\end{eqnarray}
where $\upepsilon\equiv E^\pm/\omega$, 
$\kappa=g/\omega$ reflects the coupling strength, and
$\Delta=\mu/\omega=\omega_0/(2\omega)$ \cite{AMep}. 
The Hilbert space ${\cal B}=\mathfrak{b} \otimes \mathbb{C}^2$
can be thus written as a direct sum 
${\cal B}={\cal B}_+\oplus {\cal B}_-$ of the parity eigenspaces.
The case of a displaced harmonic oscillator, which is the 
exactly solvable limit of $\hat{H}_R$ for $\mu=0$, corresponds to
$\Delta=0$, whereby the recurrence (\ref{rbmb2}) reduces 
to Eq. (A.17) of Ref. \cite{Schw}.
Because the recurrence (\ref{rbmb2}) satisfies 
the conditions that guarantee uniqueness
of the minimal solution,
i.e. each $\upvarphi^\pm(z)$ generated by 
the respective minimal solutions 
is {\em unique}, the spectrum in each parity eigenspace 
${\cal B}_\pm$ is necessarily 
{\em nondegenerate} (cf. Sec. \ref{sc:dgnr}). 
\begin{figure}
\begin{center}
\includegraphics[width=\columnwidth,clip=0,angle=0]{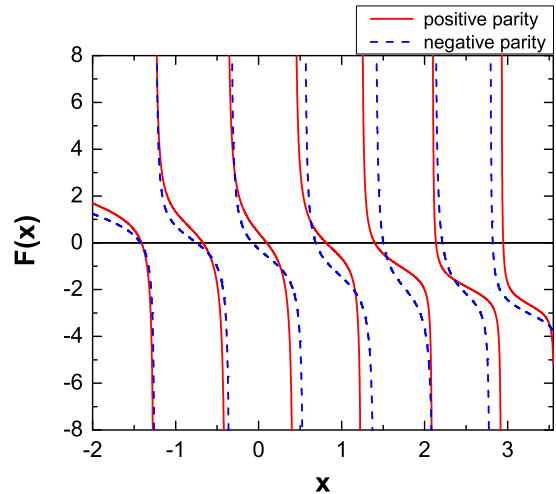}
\end{center}
\caption{$F(x)$ corresponding to the recurrence (\ref{rbmb2})
of the Rabi model in the deep strong coupling regime for $\kappa=1.4$, 
$\Delta=0.4$, $\omega=1$, and $x=\upepsilon/\kappa=E/g$. 
Zeros of $F(x)$ determine the spectrum of the Rabi model in the
respective parity subspaces ${\cal B}_\pm$.
Parity is classified according to the sign of the
eigenvalues of the parity operator $\hat{\Pi}_{FG}$.}
\label{fgrbg1p4}
\end{figure}

In what follows, section \ref{sc:ovrs} provides an 
overview of our main results.
The results are proven in the forthcoming section \ref{sc:rs},
which is divided into two subsections. 
First, subsection \ref{sc:rsfn} deal with the case 
of an arbitrary large but finite $n=N$.
The limit $N\rightarrow\infty$ is then considered 
in subsection \ref{sc:rsinfn}.
Section \ref{sc:exmp} illustrates some of 
our findings on the exactly solvable 
case of the displaced harmonic oscillator. 
In Sec. \ref{sec:disc} our results are
then extensively discussed from various angles.
Subsection \ref{sc:brk} gives
a comparison of the properties of our $F$ 
with those of Braak's functions $G_\pm$ and critically
examines his integrability arguments.
Subsection \ref{sc:dgnr} shows on a number of examples
that the present approach is a powerful 
alternative to the Frobenius analysis \cite{Frb}. 
Compared to the latter, it enables 
one an immediate insight regarding 
the nondegeneracy of the spectrum simply 
by checking that the conditions which guarantee uniqueness
of the minimal solution are satisfied.
In subsection \ref{sec:slvb} 
recent claims regarding solvability of 
the Rabi model \cite{Br,Sln} are critically examined.
A relation between the zeros of $\phi_n$ and the spectrum
is discussed in subsection \ref{sc:phn}.
Compatibility of our results with 
some other results of the theory of infinite 
continued fractions and complex analysis is demonstrated
in subsection \ref{sc:cmb}. Subsection \ref{sc:open} gives 
an overview of some open problems. 
We then conclude with Sec. \ref{sec:conc}.
Some additional technical remarks are relegated to
Appendix \ref{app:tchr}.

\section{Overview of the main results}
\label{sc:ovrs}
In the case of the Rabi model, and its special case
of the displaced harmonic oscillator, it was observed 
that the plots of $F(\upepsilon)$ corresponding to (\ref{rbmb2}) displayed 
a series of {\em discontinuous} branches 
{\em monotonically decreasing} between $+\infty$ and $-\infty$
(see Fig. \ref{fgrbg1p4} and Figs. 1,2 of Ref. \cite{AMep}).
In the present work the latter property will be proven.
First we show that $F(x)$, considered as a function 
of $x\equiv\upepsilon/\kappa=E^\pm/g$, can be 
alternatively expressed as the limit of rational functions
\begin{equation}
F(x)\equiv \lim_{n\rightarrow\infty} F_n(x) =
        a_0 + \lim_{n\rightarrow\infty} \frac{P_{n-1}^{(1)}(x)}{P_n(x)},
\label{fdfp}
\end{equation}
where $\{P_n(x)\}$ 
and $\{P_n^{(1)}(x)\}$ are associated 
systems of {\em monic orthogonal polynomials}.
(Monic means here that the coefficient of the highest
power of $x$ is one.) For $n\ge 1$, the polynomials of each  
orthogonal polynomial system (OPS) 
$\{P_n(x)\}$ and $\{P_n^{(1)}(x)\}$

\begin{itemize}

\item have {\em real} 
and {\em simple} zeros, and 

\item the zeros of $P_{n-1}^{(1)}(x)$ 
and $P_n(x)$ are {\em interlaced}.

\end{itemize}
Specifically, denote the zeros of $P_n(x)$ with degree $P_n = n$ by
$x_{n1}<x_{n2}<\ldots< x_{nn}$ and the zeros of $P_{n-1}^{(1)}(x)$ 
with degree $n-1$ by
$x_{n-1,1}^{(1)}<x_{n-1,2}^{(1)}<\ldots< x_{n-1,n-1}^{(1)}$.
Then for any $k=1,2,\ldots, n-1$
\begin{eqnarray}
x_{n,k}^{(\alpha)}<x_{n-1,k}^{(\alpha)}<x_{n,k+1}^{(\alpha)},
\label{1p5p4}
\\
x_{nk}<x_{n-1,k}^{(1)}<x_{n,k+1},
\label{3p4p1}
\end{eqnarray}
where $\alpha=0,1$ (for the sake of notation the superscript $(0)$
for $\alpha=0$ will be suppressed in what follows).
For each fixed $k$, $\{x_{nk}\}_{n=k}^\infty$ is a {\em decreasing} 
sequence and the limit 
\begin{equation}
\xi_k= \lim_{n\rightarrow\infty} x_{nk}
\label{1p5p6}
\end{equation}
exists. 
Additionally, for any finite $n$ the 
ratio in (\ref{fdfp}), also known as a {\em convergent}, enables 
a partial fraction decomposition (PFD)
\begin{equation}
\frac{P_{n-1}^{(1)}(x)}{P_n(x)}=\sum_{k=1}^n
\frac{M_{nk}}{x-x_{nk}}\cdot
\label{pfdr}
\end{equation}
The numbers $M_{nk}$ are all {\em positive}, $M_{nk}>0$,
and satisfy the condition
\begin{equation}
\sum_{k=1}^n M_{nk}=1.
\label{1p6p2}
\end{equation}
Each number $M_{nk}$ can be shown to correspond 
to the weight corresponding to the 
zero $x_{nk}$ in the Gauss quadrature formula 
for the positive definite moment functional 
${\cal L}$ associated with the OPS
 $\{P_n(x)\}$.
In the case of the displaced harmonic oscillator
and the Rabi model,
\begin{eqnarray}
M_{nk} &=&-\frac{(n+1)!}{P_{n+1}(x_{nk})P_n'(x_{nk})}
\nonumber\\
&=&\left(\sum_{l=0}^{n-1}\frac{P_l^2(x_{nk})}{(l+1)!}\right)^{-1}>0.
\label{Mnk}
\end{eqnarray}
From (\ref{pfdr}) one finds immediately that 
whenever the derivative $dF_n (x)/dx$ exists, then
\begin{equation}
\frac{dF_n(x)}{dx} <0.
\label{dfdc}
\end{equation}
Consequently, between any two subsequent $x_{nk}<x_{n,k+1}$,
where $F_n(x)$ decreases from $+\infty$ to $-\infty$,
there is exactly one zero of $F_n(x)$, in agreement 
with Fig. \ref{fgrbg1p4} and Figs. 1,2 of Ref. \cite{AMep}. 
$F_n(x)$ has its zeros and poles interlaced on the real axis.
Now the coefficient 
\begin{equation}
a_0=-x \pm (\Delta/\kappa)
\label{a0f}
\end{equation}
is {\em nonsingular}. 
Because for $x > x_0= \pm (\Delta/\kappa)$ one has $a_0<0$, 
the PFD (\ref{pfdr}) and Eq. (\ref{3p4p1}) imply that 
any two subsequent zeros $\{Z_l\}$ of $F_n(x)$ are interlaced 
for $Z_l>x_0$  as follows 
\begin{equation}
x_{nl}< Z_{l}< x_{n-1,l}^{(1)} < x_{n,l+1} < Z_{l+1}.
\label{zch1}
\end{equation}
For the zeros $Z_l\le x_0$ one has $a_0>0$ and
\begin{equation}
Z_0 < x_{n1}< x_{n-1,1}^{(1)} < Z_{1} < x_{n,2} < x_{n-1,2}^{(1)} < Z_{2} \ldots
\label{zch2}
\end{equation}
At the crossover from positive to negative $a_0$ then 
\begin{equation}
x_{n-1,l_0}^{(1)} < Z_{l_0} < x_{n,l_0+1}< Z_{l_0+1}< x_{n-1,l_0+1}^{(1)}
\label{zch3}
\end{equation}
for some $l_0$.
Thereby the above sharp inequalities 
prevent any accumulation point of the spectrum. 
That would also conclude any numerical method of computing $F(x)$ through 
Eq. (\ref{fdfp}), because of an unavoidable cutoff at some 
$n=N\gg 1$.

The above conclusions remain valid also 
in the limit $n\rightarrow\infty$. 
A point of crucial importance is that the inequality (\ref{1p5p4})
survives the limit as the sharp inequality
\begin{equation}
\xi_k<\xi_{k+1}
\label{zing}
\end{equation}
for all $k\ge 1$.
The sequence in Eq. (\ref{fdfp}) converges 
to a {\em Mittag-Leffler} PFD,
\begin{equation}
F(z) = a_0 + \sum_{k=1}^\infty \frac{{\cal M}_k}{z-\xi_k},
\label{rpthml}
\end{equation}
defining a meromorphic function
in the complex plane $\mathbb{C}$ with {\em real simple} poles 
and {\em positive} residues
\begin{equation}
0< {\cal M}_k = 
\left[\sum_{l=0}^\infty \frac{P_l^2(\xi_k)}{(l+1)!}\right]^{-1}.
\label{rsfrm}
\end{equation}
The series is absolutely and uniformly 
convergent in any finite domain having a finite distance
from the simple poles $\xi_j$, 
and it defines there a holomorphic function of $z$
[$z\in\mathbb{C}$ here and below has no relation 
to $z$ in Eq. (\ref{pss})].
One obtains at once that
\begin{equation}
\frac{dF(x)}{dx} < 0.
\label{dfdcl}
\end{equation}
Because $F(x)$ monotonically decreases from $+\infty$ to $-\infty$ 
between any its subsequent poles $\xi_k$ and $\xi_{k+1}$,
there is exactly one zero of $F(x)$ for $x\in (\xi_k,\xi_{k+1})$.
As a byproduct, the spectrum 
in each parity eigenspace ${\cal B}_\pm$ 
does not have any accumulation point.
Eventually, the knowledge of another OPS, 
$\{P_n^{(-1)}(x)\}$, enables one to determine
the expansion coefficients  of 
a physical state described by Eq. (\ref{pss}) as 
\begin{equation}
\phi_n(x)=\frac{P_{n}^{(-1)}(x)}{n!}\cdot
\label{pnPn}
\end{equation}

\section{Proof of the main results}
\label{sc:rs}
According to the Wallis formulas 
(Eqs. (III.2.1) of Ref. \cite{Chi};
Eqs. (4.2-3)  of Ref. \cite{Gt}), 
given a three-term recurrence (\ref{3trg}), 
the infinite continued fraction
in Eq. (\ref{fdfs}) can be recast as the limit 
\begin{equation}
r_0=\lim_{n\rightarrow\infty} \frac{A_n}{B_n}\cdot
\label{r0l}
\end{equation}
Here the $n$th {\em partial numerator} $A_n$ and 
the $n$th {\em partial denominator} $B_n$ are 
determined as linearly independent 
solutions of the recurrence
\begin{eqnarray}
A_n  &=& a_n A_{n-1}- b_n A_{n-2},
\label{wlfl1}\\
B_n  &=& a_n B_{n-1}- b_n B_{n-2},
\label{wlfl2}
\end{eqnarray}
where $n\ge 1$.
The $A_n$'s and $B_n$'s 
are differentiated by the initial conditions:
\begin{equation}
A_{-1}=1,~~~~A_0=0,~~~~B_{-1}=0,~~~~B_0=1.
\end{equation}
In an intriguing and peculiar world of infinite continued
fractions, the respective recurrences 
(\ref{wlfl1}) and (\ref{wlfl2})
are essentially identical 
to the initial three-term recurrence (\ref{3trg}) 
(up to the change $a_n\rightarrow -a_n$ and 
the omission of the $n=0$ term). 
For the Rabi model we have 
(Eq. (37) of Ref. \cite{AMep}, or Eq. (\ref{rbmb2})
herein above)
\begin{eqnarray}
a_n &=&-\frac{1}{(n+1)}\,
      \left(\frac{\upepsilon}{\kappa} - \bar{c}_{n}\right),~~~~~
b_n=\frac{1}{n+1},
\label{anbn}
\\
\bar{c}_{n} &\equiv& \frac{1}{\kappa }\,[n  \pm (-1)^n\Delta].
\label{cn}
\end{eqnarray}

\subsection{Arbitrary large but finite $N$}
\label{sc:rsfn}
In order to prove the properties of $F_n(x)$
defined by Eq. (\ref{fdfp}) for an arbitrary $n$, 
together with the properties listed
below, it is sufficient to prove that each 
of the recurrences (\ref{rbmb2}),
(\ref{wlfl1}), (\ref{wlfl2}) can be transformed 
into a recurrence of the type
\begin{eqnarray}
P_n(x) &=&(x-c_{n})P_{n-1}(x)- \lambda_{n} P_{n-2}(x),
\label{chi3tr}
\\
P_{-1}(x) &=&  0,~~~~~~ P_{0}(x)=1,
\label{chi3tric}
\end{eqnarray}
where $n\ge 1$, the coefficients $c_n$ and $\lambda_n$ 
are {\em real} and independent
of $x$, and $\lambda_n > 0$.
Obviously, the above recurrence defines 
a family of polynomials $\{P_n\}$ with
degree $P_n= n$. According to Favard-Shohat-Natanson theorems 
(given as Theorems I-4.1 and I-4.4 of Ref. \cite{Chi}),
satisfying the above recurrence is a necessary 
and sufficient condition that
there exists a unique positive definite moment functional
${\cal L}$, such that
for the family of polynomials $\{P_n\}$ holds
\begin{equation}
{\cal L}[1]=\lambda_1,~~~~~
  {\cal L}[P_m(x)P_n(x)]= \lambda_1\lambda_2\ldots \lambda_{n+1}\delta_{mn},
\end{equation}
$m,n=0,1,2,\ldots$ and $\delta_{mn}$ is the Kronecker symbol. 
Thereby the polynomials
$\{P_n\}$ form an OPS \cite{genr}. Because $\lambda_n > 0$, the norm 
of the polynomials $P_n$ is positive definite, 
${\cal L}[P_n^2(x)]>0$, and ${\cal L}$ 
is {\em positive definite} moment functional (p. 16 of Ref. \cite{Chi}). 

With $a_n$ and $b_n$ as in Eqs. (\ref{anbn}), 
the substitution $B_n=(-1)^n P_n/(n+1)!$ 
transforms the three-term recurrence (\ref{wlfl2})
into the recurrence of the type (\ref{chi3tr}) and (\ref{chi3tric}) with
\begin{equation}
c_n=\bar{c}_{n},
~~~~~
\lambda_n=\bar{\lambda}_n\equiv n,
\label{alr}
\end{equation}
where $\bar{c}_{n}$ has been defined by
Eq. (\ref{cn}) and $x=\upepsilon/\kappa=E/g$.
A similar substitution $A_n=(-1)^n S_{n}/(n+1)!$ transforms
(\ref{wlfl1}) into the recurrence 
\begin{equation}
S_n(x) = (x-\bar{c}_{n}) S_{n-1}- \lambda_n S_{n-2}
\label{snr}
\end{equation}
with $\lambda_n=n$, but with a ``wrong" initial condition
\begin{equation}
S_{-1}=-1,~~~~S_0=0.
\label{snic}
\end{equation}
The latter is not of the required type (\ref{chi3tric}).
Note that a recurrence of the type (\ref{snr}) yields
\begin{eqnarray}
S_1 &=& \lambda_1,
\nonumber\\
S_2 &=& (x-\bar{c}_2)\lambda_1,
\nonumber\\
S_3 &=& (x-\bar{c}_3)(x-\bar{c}_2)\lambda_1 + \lambda_3\lambda_1,
\nonumber\\
S_3 &=& (x-\bar{c}_4)(x-\bar{c}_3)(x-\bar{c}_2)\lambda_1
               +(x-\bar{c}_4)\lambda_3\lambda_1
\nonumber\\
    &&~~~~~~ +(x-\bar{c}_2) \lambda_4\lambda_1,
\nonumber\\
    && \ldots
\end{eqnarray}
Therefore, a further substitution $S_n=\lambda_1 Q_{n-1}$ transforms 
the recurrence (\ref{snr}) into
\begin{equation}
Q_{n}(x)= (x-\bar{c}_{n+1}) Q_{n-1}(x) - \lambda_{n+1} Q_{n-2}(x), 
\label{qnr}
\end{equation}
where $n\ge 1$, with the ``correct" initial conditions
\begin{equation}
Q_{-1}=0,~~~~Q_0=1.
\label{qnic}
\end{equation}
The recurrence (\ref{qnr}) together with the initial conditions
is now of the type (\ref{chi3tr}) and (\ref{chi3tric}) with
\begin{equation}
c_{n}=\bar{c}_{n+1},
~~~~~
\lambda_n=\bar{\lambda}_{n+1}=n+1.
\label{alp1r}
\end{equation}

Eventually, the substitution $\phi_n\rightarrow\bar{\phi}_n/n!$
transforms the initial recurrence (\ref{rbmb2}) into
\begin{equation}
 \bar{\phi}_{n+1}^\pm  = (x -\bar{c}_{n}) \bar{\phi}_{n}^\pm  
                        -\bar{\lambda}_n\, \bar{\phi}_{n-1}^\pm.
\label{rbmb2h}
\end{equation}
The recurrence for $\bar{\phi}_{n}^\pm$ 
is again of the type (\ref{chi3tr}) and (\ref{chi3tric}) with 
\begin{equation}
c_{n}=\bar{c}_{n-1},
~~~
\lambda_n=\bar{\lambda}_{n-1}=n-1,
\label{alm1r}
\end{equation}
where we set $\lambda_1=\bar{\lambda}_0=1\ne 0$ for $n=1$. 
Note in passing that $\lambda_1$ enters the 
recurrence (\ref{chi3tr}) only in the product $\lambda_1P_{-1}$, 
where $P_{-1}$ satisfies the initial condition (\ref{chi3tric}).
Therefore we have the freedom to set $\lambda_1$ at our will.
The initial conditions (\ref{chi3tric}) in the case
of the recurrence (\ref{rbmb2h}) for $\bar{\phi}_{n}^\pm$
are justified, because the logarithmic derivative of 
the entire function $\upvarphi(z)$ generated by the 
minimal solution (of the $n\ge 1$ part) 
of (\ref{3trg}) satisfies the {\em boundary condition} (\ref{bcc}).
Combined with the fact that in the case of the recurrence (\ref{rbmb2})
the coefficient $a_0$ is {\em nonsingular} 
[cf. Eq. (\ref{a0f})], one has necessary $\bar{\phi}_{0}^\pm \ne 0$.
A suitable rescaling, which can be absorbed into an overall
normalization prefactor, then always achieves $\bar{\phi}_{0}^\pm = 1$.

Now the respective recurrences 
for $\bar{\phi}_{n}$'s, $P_n$'s, and $Q_{n}$'s 
have all been shown to be of the type
\begin{eqnarray}
P_n^{(\alpha)}(x) &=&(x-\bar{c}_{n+\alpha})P_{n-1}^{(\alpha)}(x)
            - \bar{\lambda}_{n+\alpha} P_{n-2}^{(\alpha)}(x),
\label{chi3tra}
\\
P_{-1}^{(\alpha)}(x) &=&  0,~~~~~~ P_{0}^{(\alpha)}(x)=1,
\label{chi3trica}
\end{eqnarray}
where the coefficients $\bar{c}_n$ 
and $\bar{\lambda}_{n+\alpha}$ are {\em real} and independent
of $x$, and $\bar{\lambda}_{n+\alpha} > 0$ for $n\ge 1$.
One has $\alpha=-1,0,1$ for 
$\bar{\phi}_{n}$'s, $P_n$'s, and $Q_{n}$'s, respectively.
We continue to denote the polynomials of the OPS
for $\alpha=0$ by $\{P_n\}$.
They determine the denominators $B_n$'s in Eq. (\ref{r0l}).
The respective monic OPS with $\alpha=-1,1$ are called 
{\em associated} to $P_n$'s and will be denoted by 
$\{P_n^{(\alpha)}\}$ (see Sec. III-4 of Ref. \cite{Chi}).
Because 
\begin{equation}
A_n=\frac{(-1)^n P_{n-1}^{(1)}}{(n+1)!},~~~~~B_n=\frac{(-1)^n P_n}{(n+1)!},
\end{equation}
it follows at once that the ratio (\ref{r0l}) can be expressed as
the limit of the ratios of the orthogonal monic polynomials
\begin{equation}
r_0=\lim_{n\rightarrow\infty} \frac{P_{n-1}^{(1)}(x)}{P_n(x)}\cdot
\label{r0lo}
\end{equation}
The properties listed below Eq. (\ref{fdfp}) 
follow straightforwardly 
from the classic theory of orthogonal polynomials 
(see esp. Secs. I.4-6
and III.1-4 of Ref. \cite{Chi}). 
The zeros of the polynomials of any OPS 
are {\em real} and {\em simple} (Theorem I-5.2 of Ref. \cite{Chi}). 
Furthermore, the zeros of any two subsequent 
polynomials $P_n(x)$ and $P_{n+1}(x)$ of an OPS
mutually separate each other (Theorem I-5.3 of Ref. \cite{Chi}).
The separation property of zeros (\ref{3p4p1}) follows
from Theorem III-4.1 of Ref. \cite{Chi}.
Eqs. (\ref{1p5p6}) and 
(\ref{1p6p2}) follow from Eqs. (I-5.6) and (I-6.2) 
of Ref. \cite{Chi}, where we have assumed ${\cal L}[1]\equiv \mu_0=1$.
The partial fraction decomposition (\ref{pfdr}) follows from
Theorem III-4.3 of Ref. \cite{Chi}.
The positivity of $M_{nk}$ in Eq. (\ref{Mnk}) follows from the
Christoffel-Darboux identity (Eq. (I-4.13) of \cite{Chi}),
\begin{equation}
P_{n+1}'(x)P_n(x) - P_n'(x)P_{n+1}(x) > 0,
\label{cdid}
\end{equation}
which for $x=x_{nk}$ reduces to
\begin{equation}
P_{n+1}(x_{nk}) P_n'(x_{nk}) < 0,
\label{cdidsc}
\end{equation}
where the prime denotes derivative.
The 2nd of Eqs. (\ref{Mnk}) follows 
from Theorem I-4.6 of Ref. \cite{Chi}.
Thereby our results for any finite $n=N$ have been proved.

Note in passing that, in virtue of the Gauss
quadrature formula (Eq. (II-3.1) of Ref. \cite{Chi}),
the coefficients $M_{nk}$'s in the PFD (\ref{pfdr}) satisfy
\begin{equation}
\sum_{k=1}^n M_{nk}x_{nk}^l=\mu_l~~~~~~~(l=0,1,2,\ldots, 2n-1),
\end{equation}
where $\mu_l$'s are the corresponding moments
of the positive definite moment functional, ${\cal L}[x^l]=\mu_l$.
(The positivity of ${\cal L}$ implies $\mu_{2l}>0$, but not necessarily
$\mu_{2l+1}>0$.)

\subsection{The limit $N\rightarrow\infty$}
\label{sc:rsinfn}
According to the representation theorem (Theorem II-3.1 of Ref. \cite{Chi}), 
the weight function $\psi$ of the positive moment functional 
${\cal L}$ (also called {\em distribution} function \cite{Chi}),
\begin{equation}
{\cal L}[x^n]=\int_{-\infty}^\infty x^n\, d\psi(x)
                       =\mu_n ~~~~~~~~(n=0,1,\ldots),
\label{mf}
\end{equation}
is the limit of a sequence of bounded, right continuous, 
nondecreasing step functions $\psi_n(x)$'s, 
\begin{eqnarray}
\psi_n(x) &=& 0~~~~~~~~~ (-\infty\le x < x_{n1}), 
\nonumber\\
\psi_n(x)&=& M_{n1}+\ldots+ M_{np} ~~(x_{np} \le x < x_{n,p+1}), 
\nonumber\\
\psi_n(x)&=& \mu_0~~~~~~~ (x \ge x_{nn}) .
\label{psin}
\end{eqnarray}
Consequently
\begin{itemize}

\item $\psi_n(x)$ has exactly $n$ points 
of increase, $x_{nk}$,

\item the discontinuity of $\psi_n(x)$ at each $x_{nk}$  
equals $M_{nk}$ ($k = 1, 2, \ldots, n$),

\item at least the first $(2n-1)$ moments of the weight 
function $\psi_n(x)$ are identical with those of $\psi(x)$, i.e.,
\begin{equation}
\int_{-\infty}^\infty x^l\,d\psi_n(x)=\mu_l 
~~~~~~~~~~(l = 0,1, 2, \ldots, 2n-1).
\end{equation}

\end{itemize}
Obviously, for any $z\in\mathbb{C}$ different
from the zeros $x_{nk}$'s the PFD in Eq. (\ref{pfdr}) 
can be expressed as
\begin{equation}
\frac{P_{n-1}^{(1)}(z)}{P_n(z)} 
   =\sum_{k=1}^n \frac{M_{nk}}{z-x_{nk}}
     =\int_{-\infty}^\infty \frac{d\psi_n(x)}{z-x}\cdot
\label{pfdrl}
\end{equation}
According to Hamburger's Theorem XII' \cite{Hb1}, 
the function
\begin{equation}
f(z) = \int_{-\infty}^\infty \frac{d\psi(x)}{z-x},
\label{fzd}
\end{equation}
where the Stieltjes integral measure $d\psi$ has been defined 
through the limit of $\psi_n(x)$'s,
is a {\em regular analytic} function in any closed finite region $\Omega$ 
of the complex plane which does not contain any part of the real axis.
The convergents in Eq. (\ref{pfdrl}) 
converge {\em uniformly} to $f(z)$ in $\Omega$.
According to Definition III-1.1 of \cite{Chi}, 
the infinite continued fraction in Eqs. (\ref{fdfs}) 
and (\ref{r0l}) then converges and
\begin{equation}
F(z) = a_0 + \int_{-\infty}^\infty \frac{d\psi(x)}{z-x}\cdot
\label{rpth}
\end{equation}

So far we have mostly summarized the relevant classical results
of Hamburger \cite{Hb1}. A point of crucial importance in our case is that
the resulting Stieltjes measure $d\psi(x) \equiv \psi(x) - \psi(x-0)$
is necessarily {\em discrete}. 
(Here $\psi(x-0)$ denotes the left-side limit of $\psi$
at $x$, $\psi(x-0)=\lim_{x_n\rightarrow x,\, x_n<x} \psi(x_n)$.)
To this end, we first show that the set of zeros 
$x_{nk}$ extends beyond any 
bound up to at $+\infty$. Denote 
\begin{equation}
\sigma \equiv \lim_{j\rightarrow \infty} \xi_j,
\label{sgdf}
\end{equation}
where $\xi_j$'s are the limit zero points defined by 
Eq. (\ref{1p5p6}).
According to Eq. (IV-3.7) of Ref. \cite{Chi},
a sufficient condition for $\sigma=\infty$
is that
\begin{equation}
\lim_{n\rightarrow \infty} c_n=\infty~~~ \mbox{and}~~~
\lim \sup_{n\rightarrow \infty} 
    \frac{\lambda_{n+1}}{c_n c_{n+1}}<\frac{1}{4}\cdot
\label{sginf}
\end{equation}
In the present case of the Rabi model, with $\bar{c}_{n}$ and 
$\bar{\lambda}_n$ defined by Eq. (\ref{alr}), one has 
$\bar{\lambda}_{n+1}/(\bar{c}_n \bar{c}_{n+1})={\cal O} (n^{-1})$.
The conditions (\ref{sginf}) are then obviously satisfied,
\begin{equation}
\lim_{n\rightarrow \infty} \bar{c}_{n}=\infty,~~~~~~
\lim \sup_{n\rightarrow \infty} 
     \frac{\bar{\lambda}_{n+1}}{\bar{c}_n \bar{c}_{n+1}}=0.
\label{sginfrm}
\end{equation}
Now the condition $\sigma=\infty$ ensures that the limit zero points 
$\xi_k$ defined by (\ref{1p5p6}) are all {\em distinct}, i.e., 
Eq. (\ref{zing}) holds. 
Indeed, if $\xi_k=\xi_{k+1}$ for some $k$, then 
$\xi_k$ is a limit point of $\xi_l$'s (Theorem II-4.4 of Ref. \cite{Chi}).
According to Theorem II-4.6 of Ref. \cite{Chi},
if $\xi_k=\xi_{k+1}$ for some $k\ge 1$, then 
\begin{equation}
\xi_k= \sigma \equiv \lim_{j\rightarrow \infty} \xi_j.
\end{equation} 
(Such a separation of the limit zero points $\xi_k^{(\alpha)}$'s
applies also to the other two OPS with $\alpha\ne 0$.)
Because of the sharp inequality (\ref{zing}),
the Stieltjes weight function $\psi$ 
satisfies $\psi=$ const on any interval $x\in(\xi_k,\xi_{k+1})$.
Consequently $d\psi$ is a {\em discrete} measure.
Moreover, the measure 
$d\psi$ is unambiguously determined
(see footnote 50 on p. 268 of Ref. \cite{Hb1}).
The determinacy of $d\psi$, and the Stieltjes weight function $\psi$ 
[assuming the normalization $\psi(-\infty)=0$],
follows also independently from Carleman's 
criterion which says that the moment problem is determined if  
(cf. Eq. (VI-1.14) of Ref. \cite{Chi})
\begin{equation}
\sum_{l=1}^\infty \lambda_l^{-1/2} =\infty.
\label{crlcr}
\end{equation}
The latter is obviously satisfied in our case.

Now the support of the measure induced by $\psi$, 
or briefly the spectrum of $\psi$, is precisely the set of 
all $\xi_k$'s (pp. 63 and 113 of Ref. \cite{Chi}).
Therefore, by the very definition of the spectrum of $\psi$
(p. 51 of Ref. \cite{Chi}),
the residues in (\ref{rpth}) are strictly positive,
$0< d\psi(\xi_k) = {\cal M}_k$. This proves Eq. (\ref{rsfrm})
for $z\in\Omega$ having a finite distance $\delta>0$ from the real axis.
As the result, the Stieltjes integrals 
in Eqs. (\ref{fzd}) and (\ref{rpth}) reduce to infinite sums. 

\begin{figure}
\begin{center}
\includegraphics[width=\columnwidth]{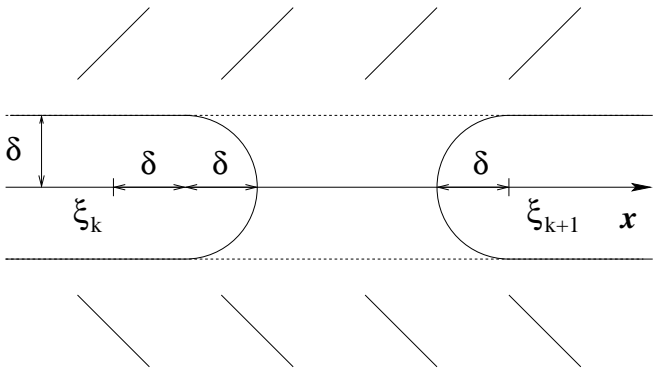}
\end{center}
\caption{Analyticity domain of $F(z)$ guaranteed by 
Hamburger's theorem \cite{Hb1} extends to the half-planes
above and below the real axis at the distance of at least 
$\delta$ from the latter as showed
by the dashed lines. 
In the case of the Rabi model there is possible to extend
the domain of analyticity of $F(z)$ through the real axis in any
interval $(\xi_{k},\xi_{k+1})$ and to connect the upper and lower
half-planes.}
\label{fggand}
\end{figure}

In what follows we show by the Stieltjes-Vitali theorem 
(cf. p. 121 of Ref. \cite{Chi}; p. 144 of Ref. \cite{Gr})
that the convergents in Eq. (\ref{pfdrl}) 
converge {\em uniformly} to 
a {\em regular analytic} function $f(z)$ represented by Eq. (\ref{fzd})
not only in any closed finite $\Omega\cap\mathbb{R}=\emptyset$, such 
as in Hamburger's Theorem XII' \cite{Hb1},
but also on the real axis for 
$z=x\in (\xi_{k}+2\delta,\xi_{k+1}-\delta)$, where
$\delta>0$ is sufficiently small real number
(see Fig. \ref{fggand}). To this end 
we remind that the sum over all $M_{nk}$'s satisfies Eq. (\ref{1p6p2}).
Therefore (cf. Eq. (56) of Ref. \cite{Gr}),
\begin{equation}
\left|\frac{P_{n-1}^{(1)}(z)}{P_n(z)} \right| 
           \le \sum_{k=1}^n \frac{M_{nk}}{|z-x_{nk}|}
\le  \frac{1}{\delta},
\label{pfdrlll}
\end{equation}
where $\delta=\min |z-x_{nk}|$. 
Now select any pair of subsequent 
limit zero points $\xi_{k}$ and $\xi_{k+1}$. 
Because the limit zero points are separated, there exists 
some infinitesimal $\delta>0$
so that $\xi_{k}+2\delta<\xi_{k+1}-\delta$. Now consider 
$z=x\in (\xi_{k}+2\delta,\xi_{k+1}-\delta)$.
Because each sequence $x_{nk}$ is decreasing 
with increasing $n$, $x_{nk} \notin (\xi_{k}+\delta,\xi_{k+1})$ 
for sufficiently large $n\ge N$.
Then $\min |x-x_{nk}|\ge \delta>0$. This establishes 
a uniform bound on the convergents
for $z=x\in (\xi_{k}+2\delta,\xi_{k+1}-\delta)$. 
The latter bound can be obviously extended to any closed
region in the complex plane bounded with a semicircle of radius $\delta$ at  
$\xi_{k}+\delta$ and a semicircle of radius $\delta$ at $\xi_{k+1}$ and 
having a distance at least $\delta$ from the real axis 
for Re $z\in (-\infty,\xi_{k}+\delta]$ and Re $z\in [\xi_{k+1},\infty)$
(see Fig. \ref{fggand}).
Analyticity is then established by the Stieltjes-Vitali theorem
\cite{Chi,Gr}. Thus $F(z)$
takes on the form of the {\em Mittag-Leffler} PFD (\ref{rpthml}),
which is absolutely and uniformly 
convergent in any finite domain having a finite distance
from the simple poles $\xi_k$'s.

On combining the special cases of 
Eqs. (I-4.12) and (III-4.4) of Ref. \cite{Chi}
for $x=x_{n+1,k}$ and on substituting for $P_n(x_{n+1,k})$
from the former to the latter, one obtains
\begin{equation}
\frac{P_{n}^{(1)}(x_{n+1,k})}{P_{n+1}'(x_{n+1,k})}
=
\left[\sum_{l=0}^{n} \frac{P_l^2(x_{n+1,k})}{(l+1)!}\right]^{-1}.
\label{ppbn}
\end{equation}
On comparing with the right-hand side of Eq. (\ref{rsfrm}) one finds
that at the support of $d\psi$ the left-hand side
of (\ref{ppbn}) has a {\em nonzero} limit for $n\rightarrow\infty$.
That implies the sharp inequality
\begin{equation}
\xi_{k}<\xi_{k}^{(1)}<\xi_{k+1},
\label{3p4p1l}
\end{equation}
meaning that the interlacing property 
of zeros (\ref{3p4p1}) for a finite $n$ 
survives the limit $n\rightarrow\infty$. 
Note that the sum such as in Eqs. (\ref{rsfrm}) 
and (\ref{ppbn}) enters also the celebrated Chebyshev inequalities 
(cf. Theorem II-5.5 of Ref. \cite{Chi}).
The sharp inequalities (\ref{zch1}) and (\ref{zch2})
combined with the separation of zeros in the limit $n\rightarrow\infty$
expressed by Eqs. (\ref{zing}) and (\ref{3p4p1l}) then 
prevent any accumulation point of the spectrum.

\section{Example of the displaced harmonic oscillator}
\label{sc:exmp}
The recurrence (\ref{rbmb2}) has for $\Delta= 0$, i.e., in 
the case of the displaced harmonic oscillator,  
a unique solution \cite{Schw}
\begin{equation}
\phi_n=\kappa^{-n} L_{n}^{(\upepsilon +\kappa^2-n)}(\kappa^2),
\label{cns}
\end{equation}
where $L_{n}^{(u)}$ are 
generalized Laguerre polynomials of degree $n$ \cite{Lgp}
(note different sign of $\kappa$ compared to Eq. 
(2.16) of Schweber \cite{Schw}). In order to show
explicitly that each $\phi_n$ is an
orthogonal polynomial of degree $n$ in energy parameter
$x$, one makes use of that the associated Laguerre polynomials 
are related to the Charlier polynomials 
(cf. Eq. VI-1.5 of Ref. \cite{Chi})
\begin{equation}
C_n^{(u)}(\zeta)=n! L_n^{(\zeta-n)}(u).
\label{chrp}
\end{equation}
Therefore,
\begin{equation}
n!\phi_n=P_n^{(-1)}(x) 
      = \kappa^{-n} C_n^{(\kappa^2)}(\zeta),
\label{pnm1}
\end{equation}
where $\zeta=\upepsilon +\kappa^2=\kappa x +\kappa^2$.
On substituting explicit form of
the Charlier polynomials (cf. Eq. VI-1.2 of Ref. \cite{Chi}),
\begin{equation}
\phi_n = \sum_{j=0}^n (-1)^{n-j}
\frac{\kappa^{n-2j}}{(n-j)!j!}\, \prod_{k=0}^{j-1} (\zeta-k).
\label{cnsp}
\end{equation}
The eigenvalues of the displaced harmonic oscillator \cite{Schw}
\begin{equation}
\upepsilon_l=l-\kappa^2
\label{bslnc}
\end{equation}  
correspond to $\zeta=l\in \mathbb{N}$ (including $l=0$).
Obviously, each $\phi_n$ is also a 
polynomial of degree $n$ in energy parameter $x$:
\begin{eqnarray}
\phi_0 &=& L_{0}^{(\zeta)}(\kappa^2) =C_0^{(\kappa^2)}(\zeta)=1,
\nonumber\\
\phi_1 &=& \kappa^{-1} L_{1}^{(\zeta-1)}(\kappa^2)
=\kappa^{-1} C_1^{(\kappa^2)}(\zeta) =\upepsilon/ \kappa=x,
\nonumber\\
\phi_2 &=& \kappa^{-2} 
\left[\frac{\kappa^4}{2} -\zeta \kappa^2 
      + \frac{\zeta(\zeta-1)}{2}\right]
\nonumber\\
&=& 
 \frac{x}{2!}\, \left(x-\frac{1}{\kappa}\right) -\frac{1}{2!},
\nonumber\\
\phi_3 &=& \kappa^{-3}
\left[
-\frac{\kappa^6}{3!} 
+ \frac{\zeta \kappa^4}{2}  
\right.
\nonumber\\
&&
\left.
-\frac{\zeta(\zeta-1)\kappa^2}{2} 
+ \frac{\zeta(\zeta-1)(\zeta-2)}{6}\right]
\nonumber\\
&=& \frac{x}{3!}\, 
\left(x-\frac{1}{\kappa}\right)\left(x-\frac{2}{\kappa}\right) 
           -\frac{x}{2!} +\frac{1}{3\kappa}\cdot
\end{eqnarray}
Note that $\phi_1/\phi_0= x$, which 
is exactly the $n=0$ part of Eq. (\ref{rbmb2}).

Let instead of the recurrence (\ref{chi3tr}) polynomials $Q_n$ satisfy 
\begin{equation}
Q_n(x)=\left(x -\frac{c_n-q}{p}\right)\, Q_{n-1}(x)
-\frac{\lambda_n}{p^2}\, Q_{n-2}(x).
\label{qnrt}
\end{equation}
Then
\begin{equation}
Q_n(x)=p^{-n}P_n(px+q) ~~~~~(p\ne 0).
\label{qnf}
\end{equation}
Given Eqs. (\ref{pnm1}), (\ref{qnrt}), and (\ref{qnf}),
and because additionally $\lambda_0=\lambda_1=1$,
\begin{equation}
P_n(x)=P_n^{(-1)}[x-(1/\kappa)]= 
 \kappa^{-n} C_{n}^{(\kappa^2)}(\zeta-1).
\end{equation}

\section{Discussion}
\label{sec:disc}
Our recent work has been driven by the curiosity as to
what extent the recurrence coefficients 
$a_n$ and $b_n$ of a model from ${\cal R}$ 
determine the model basic properties \cite{AMep,AMcm}. 
Earlier we looked at the problem from 
the perspective of the minimal solutions \cite{Gt} of
the recurrence (\ref{3trg}) and showed that the spectrum 
of any quantum model from ${\cal R}$ can be obtained as zeros of 
a transcendental function $F$ \cite{AMep,AMcm,AMr}.
In the present work we took a complementary view and analyzed
in detail the analytic structure of Schweber's 
quantization condition \cite{Schw}.
We showed that the function $F$ originally defined
by Eqs. (\ref{fdf}) and (\ref{eth})
can alternatively be  represented
by the {\em Mittag-Leffler} PFD (\ref{rpthml})
with repelling zeros and with positive residues ${\cal M}_k$ 
defined by (\ref{rsfrm}). The latter enabled to prove
the monotonicity of $F(z)$ and that 
the spectrum of the Rabi model in each 
parity eigenspace ${\cal B}_\pm$ does not have any
accumulation point. 

We have presented our results while treating 
the special case of the Rabi model.
However, as obvious from the proof, our results remain 
valid for any model of the class ${\cal R}$
which can be reduced to
the recurrence of the type (\ref{chi3tr}) and (\ref{chi3tric})
and which satisfies the conditions (\ref{sginf}).
(More general sufficient conditions than (\ref{sginf}) can be found 
in Sec. IV-3 of Ref. \cite{Chi}.)
Essential for obtaining our results
was to start with the pair of the recently uncovered
parity resolved three-term recurrences (\ref{rbmb2}) 
(cf. Eq. (37) of Ref. \cite{AMep}).
The recurrences are different from 
the original recurrence for the Rabi model \cite{Schw}, 
\begin{equation}
\phi_{n+1} - \frac{f_n(\zeta)}{(n+1)}\,
             \phi_n + \frac{1}{n+1}\, \phi_{n-1}=0,
\label{sa8}
\end{equation}
where 
\begin{equation}
f_n(\zeta)=2\kappa+\frac{1}{2\kappa}
\left(n-\zeta - \frac{\Delta^2}{n-\zeta}\right),
\label{f-n}
\end{equation}
$\kappa=g/\omega$ and $\Delta=\mu/\omega$ are as in Eq. (\ref{rbmb2})
(cf. Eq. (A8) of Schweber \cite{Schw}, which has mistyped 
sign in front of his $b_{n-1}$, and Eqs. (4) and (5) of \cite{Br}).
The dimensionless energy parameter $\zeta=(E/\omega)+\kappa^2$ 
is the same as in Sec. \ref{sc:exmp}.

\begin{figure}
\begin{center}
\includegraphics[width=\columnwidth]{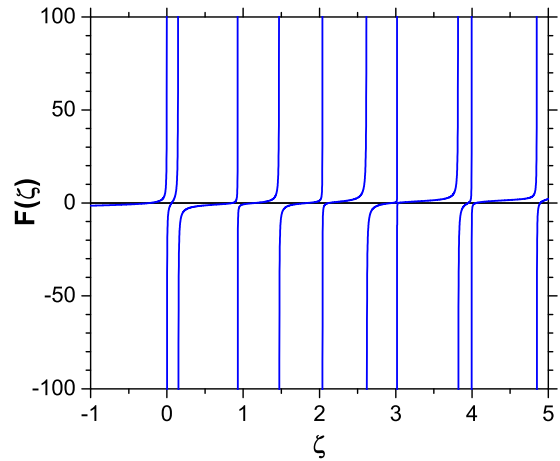}
\end{center}
\caption{\label{fgf0f} Plot of $F(\zeta)$ corresponding
to the recurrence (\ref{sa8}) of the Rabi model for 
$g=0.7$, $\Delta=0.4$, and $\omega=1$,
i.e. the same parameters as for $G_\pm(\zeta)$ in Fig. 1 of
Ref. \cite{Br}, shows corresponding 
zeros at $\approx -0.217805$, $0.0629563$,
$0.86095$, $1.1636$,  $1.85076$, etc.
}
\end{figure}

Because $f_n(\zeta)$ in Eq. (\ref{f-n}) contains $\zeta$ 
both in the numerator 
and denominator, the recurrence (\ref{sa8}) does not reduce to 
that of the type (\ref{chi3tr}) and (\ref{chi3tric}) obeyed by OPS's.
Nevertheless, as shown in Fig. \ref{fgf0f}, the 
transcendental function $F(\zeta)\equiv - f_0(\zeta)+ r_{0}$
corresponding to (\ref{sa8})
still displays a series of {\em discontinuous} branches 
{\em monotonically} extending between $-\infty$ and $+\infty$,
and the spectrum can be obtained
as zeros of $F(\zeta)$ (cf. Fig. 1 of Ref. \cite{AMcm,AMr}).
However, one can no longer guarantee that all the residues
${\cal M}_k$ are positive, nor ensure the sharp inequality (\ref{zing}).
Note that a kind of a {\em Mittag-Leffler} PFD (\ref{rpthml})
is rather general (cf. Eq. (67) of Grommer \cite{Gr}). 
It also applies to OPS's which 
distribution function has a bounded denumerable 
spectrum with a finite number of limit points 
(cf. Theorems 3.2 and 5.4 of Ref. \cite{Mk}).

\subsection{A comparison with Braak's functions and integrability}
\label{sc:brk}
In a recent letter \cite{Br}, Braak claimed to have
solved the Rabi model analytically 
(see also Viewpoint by Solano \cite{Sln}). 
He suggested \cite{Br} that a {\em regular} 
spectrum of the Rabi model in the respective
parity eigenspaces was 
given by the zeros of transcendental functions
\begin{equation}
G_\pm(\zeta)=\sum_{n=0}^\infty K_n(\zeta,\kappa)
\left[1\mp\frac{\Delta}{\zeta-n}\right]\kappa^n.
\label{sol}
\end{equation}  
Here the coefficients $K_n(\zeta,\kappa)$ were obtained recursively 
by solving the Poincar\'{e} difference equation 
(\ref{sa8}) upwardly starting from the initial condition 
\begin{equation}
K_1/K_0=f_0(\zeta) = 2\kappa-\frac{1}{2\kappa}
\left(\zeta - \frac{\Delta^2}{\zeta}\right).
\label{rbc}
\end{equation}
In general that yields the coefficients $K_n$ 
as the {\em dominant} solution
of the three-term recurrence (\ref{sa8}) \cite{AMep}.

Braak argued that between
subsequent poles of the term in the square bracket 
in (\ref{sol}) at $\zeta=n$ and $\zeta=n+1$ the function
$G_\pm(\zeta)$ takes on zero value: 
\begin{itemize}

\item {\em once} - by implicitly presuming that at one 
of the poles
$G_\pm(\zeta)$ goes to $+\infty$ and at the neighboring pole goes 
to $-\infty$, with a monotonic
behavior from $+\infty$ to $-\infty$ between the poles;

\item {\em twice} - implicitly presuming that $G_\pm(\zeta)$ goes to 
one of $\pm\infty$ at both subsequent poles 
of the term in the square bracket 
in (\ref{sol}), and in between the poles 
it has rather {\em featureless} behavior,
e.g., similar to the cord hanging on two posts;  

\item {\em none} -  occurs under the 
similar circumstances as described
in the previous item, if the ``cord is too short", e.g., it does not
stretch sufficiently up or down as to cross the abscissa.

\end{itemize}
Braak's arguments regarding integrability of the
Rabi model then rely heavily on the above properties.
However, the above behavior can be merely regarded as 
an {\em unproven} hypothesis. 
There is no proof in Ref. \cite{Br} that 
this is the only possible behavior. 
In this regard, Eqs. (\ref{sa8}) and (\ref{f-n}) show that 
$K_n(\zeta,\kappa)$ is a complicated sum 
of polynomial fractions with increasing 
polynomial degree in both the numerator and denominator. 
Any given $K_n(\zeta,\kappa)$ has its own zeros and poles structure
for it comprises all singular terms $\Delta^2/(l-\zeta)$ 
between $l=0$ and $l=n-1$, including their mutual products.
Additionally, in spite of the suggestive notation 
$K_n(\zeta)$ employed in Ref. \cite{Br}, 
any $K_n(\zeta,\kappa)$ is a rational function of 
$\kappa$ comprising terms between $\kappa^{-n}$ and 
$\kappa^n$. Indeed, the $\Delta$-independent contribution
to $K_n(\zeta,\kappa)$ is given by Eq. (\ref{cnsp}).
Consequently, (i) the representation (\ref{sol})
hides additional poles in $\zeta$ and 
(ii) it cannot be excluded that between any two 
subsequent poles of the term in the square bracket in (\ref{sol}) 
the function $G_\pm(\zeta)$ would display much more complicated
behavior than that assumed by Braak \cite{Br}.

Analogous objections apply to the proposed functional 
form of $G_\pm(\zeta)$ in Eq. (6) of Ref. \cite{Br},
\begin{equation}
G_\pm(\zeta)= G_\pm^0(\zeta) + \sum_{n=0}^\infty
\frac{h_n^\pm}{\zeta-n},
\label{bre6}
\end{equation}
where $G_\pm^0(\zeta)$ is entire in $\zeta$.
Without any control over $G_\pm^0(\zeta)$ it is impossible
to make any definite statement on the number of zeros
of $G_\pm(\zeta)$ in any predetermined interval $\zeta\in (n,n+1)$.
This should be contrasted with the Mittag-Leffler PFD (\ref{rpthml}) 
of our $F(x)$ function, which is much stronger result 
than Eq. (\ref{bre6}) of Braak \cite{Br}.
There is no entire function 
contribution in our PFD (\ref{rpthml}).
Additionally, Braak \cite{Br} cannot say anything about the residues
$h_n^\pm$, whereas  the residues ${\cal M}_k$ in 
the Mittag-Leffler PFD (\ref{rpthml}) are all positive and given 
by Eq. (\ref{rsfrm}).

\subsection{Absence of degeneracies}
\label{sc:dgnr}
In the case of a displaced harmonic oscillator, which 
is the special case
of $\hat{H}_R$ in (\ref{rabih}) for $\mu=0$,
the recurrence (\ref{3trg}) becomes (cf. 
Eq. (A.17) of Ref. \cite{Schw})
\begin{equation}
\phi_{n+1} + \frac{n-x}{(n+1)\kappa}\, \phi_{n} 
                   + \frac{1}{n+1}\, \phi_{n-1}=0.
\label{sa17}
\end{equation}
The recurrence coefficients are nonsingular.
Therefore, the uniqueness of the {\em minimal} 
solution $\{\phi_n\}_{n=0}^\infty$ of the recurrence (\ref{sa17}) 
for any value of physical parameters \cite{Schw,Gt,AMep,AMcm} implies 
a {\em unique} entire function 
$\upvarphi(z)=\sum_{n=0}^\infty \phi_n z^n$ 
in the Bargmann space $\mathfrak{b}$.
Consequently, the spectrum of the displaced harmonic oscillator
when considered in $\mathfrak{b}$, i.e. (cf. Eq. (2.1) of Ref. \cite{Schw})
\begin{equation}
\hat{H}_{dho} =
\omega \hat{a}^\dagger \hat{a}
     + \lambda (\hat{a}^\dagger + \hat{a}),
\label{dhoh}
\end{equation}
is necessarily {\em nondegenerate} with a unique eigenstate 
$\Phi(z)\equiv \upvarphi(z)\in\mathfrak{b}$ for any value of physical 
parameters.
The spectrum becomes {\em doubly-degenerate} only 
if considered as the special case of $\hat{H}_R$ in Eq. (\ref{rabih}) 
for $\mu=0$, i.e.
when described by 
\begin{equation}
\hat{H}_{dho} =
\omega \mathds{1} \hat{a}^\dagger \hat{a} 
         + \lambda\sigma_3 (\hat{a}^\dagger + \hat{a})
\label{dholh}
\end{equation}
in the product Hilbert 
space ${\cal B}=\mathfrak{b}\otimes \mathbb{C}^2$.
In the latter case, the {\em unique} $\upvarphi(z)$ 
could be substituted into Eq. (\ref{fgr}) 
to construct {\em two} different degenerate 
parity eigenstates $\Phi^\pm \in {\cal B}$.

As another example, consider the original parity unresolved  
recurrence for the Rabi model (\ref{sa8}). The latter
has the coefficients $a_n=-f_n(\zeta)/(n+1)$,
where each $f_n(x)$ given by Eq. (\ref{f-n}) has a simple pole
at $\zeta=n\in\mathbb{N}$. At the singularities, the 
conditions which guarantee uniqueness
of the minimal solution are violated.
Therefore, degeneracies of the Rabi model could only occur
at the baselines $\zeta=n\in\mathbb{N}$ (including $n=0$),
where different parity solutions considered 
as a function of energy are allowed to intersect.
Thereby the original result of Kus \cite{Ks} has been
rederived straightforwardly within our approach.

In the present case of parity resolved three-term 
recurrences for the Rabi model (\ref{rbmb2}), the recurrence
coefficients are regular 
and the recurrences (\ref{rbmb2}) satisfy 
the conditions which guarantee uniqueness
of the minimal solution for any value of physical parameters.
Therefore, the spectrum in each parity eigenspace 
${\cal B}_\pm$ is necessarily {\em nondegenerate}.

The nondegeneracy is not new result - it was initially 
obtained by Kus \cite{Ks} within the framework of Frobenius's 
analysis of regular singular points \cite{Frb}.
Yet it is both stimulating and inspiring that
our approach based on OPS's enabled us to prove
the basic analytic property of the Rabi model 
independently, without any recourse 
to its earlier proof. The above examples 
show that the present approach is a powerful 
alternative to the Frobenius analysis \cite{Frb}. 
Compared to the latter, 
it enables one to straightforwardly draw conclusions regarding 
the degeneracy or nondegeneracy of the spectrum simply 
by checking if the conditions which guarantee uniqueness
of the minimal solution are satisfied
(cf. the Poincar\'{e} theorem (Theorem 2.1 in Ref. \cite{Gt}), or
the Perron and Kreuser generalizations the Poincar\'{e} theorem
(Theorems 2.2 and 2.3(a) in Ref. \cite{Gt}).

Note in passing that the conventional recurrences 
(\ref{sa8}) (for $\zeta\notin \mathbb{N}$) and (\ref{sa17})
introduced by Schweber \cite{Schw} represent the special 
case of Poincar\'{e} recurrences. The latter is
characterized in that the respective coefficients 
$a_n$ and $b_n$ in (\ref{3trg}) have finite 
limits $\bar{a}$ and $\bar{b}$ 
\cite{Schw,Gt,AMep,AMcm}. The recurrences (\ref{sa8}) 
and (\ref{sa17}) correspond to the 
choice of $\varsigma=0$ and $\upsilon=-1$ in Eq. (\ref{rcd}), 
and yield $\tau=1$. 
They only differ in the value of 
$\bar{a}=-1/(2\kappa)$ in (\ref{sa8})
and $\bar{a}=1/\kappa$ in (\ref{sa17}), 
whereas $\bar{b}=0$ in both examples.

\subsection{Algebraic solvability}
\label{sec:slvb}
The notion of quasi-exact solvability (QES) has been 
introduced to characterize the quantum models 
possessing a {\em finite} number of eigenvalues and 
corresponding eigenfunctions that can be determined algebraically
\cite{TU,Trb,BD,KUW}. A typical quasi-exactly solvable 
Schr\"{o}dinger operator can be expressed as a polynomial 
of degree at most two in the
generators of the $sl(2,\mathbb{R})$ algebra \cite{Trb,FGR},
\begin{equation}
H= \sum_{k,m} q_{km} J_k^n J_m^n + \sum_{m} q_{m} J_m^n + q_{*},
\end{equation}
where $q_{km}$, $q_{m}$, $q_{*}$ are some 
{\em real} constants, 
\begin{equation}
J_-^n=\partial_z,~~~~~J_0^n=z\partial_z-\frac{n}{2},
              ~~~~~J_+^n=z^2\partial_z-nz,
\label{sl2r}
\end{equation}
$n$ is an integer, and
\begin{equation}
[J_+^n,J_-^n]=2J_0^n,~~~~~[J_0^n,J_\pm^n]=\pm J_\pm^n.
\label{sl2cr}
\end{equation}
The Rabi model \cite{KKT} allows for such a representation
in terms of the generators of $sl(2,\mathbb{R})$ algebra
only for the energies corresponding to the eigenvalues
of the displaced harmonic oscillator given by Eq. (\ref{bslnc})
(cf. Eq. (12) of Ref. \cite{KKT}), or equivalently for
$\zeta=l\in\mathbb{N}$. 
The baselines $\upepsilon= l-\kappa^2$ have been identified 
in the preceding section as the only place where degenerate
state could occur. This is indeed exemplified
by the Juddian exact isolated analytic solutions \cite{Jd,Ks,KL}. 
The latter correspond to the {\em degenerate} polynomial 
solutions of the Rabi model \cite{Ks}.
This shows that the very existence of the 
the parity {\em degenerate} Juddian exact isolated
analytic solutions is a direct consequence of 
the quasi-exact solvability of the Rabi model \cite{KKT}.

However, not any quasi-exactly solvable Schr\"{o}dinger operator 
can be expressed in terms of the quadratic elements
of an enveloping $sl(2,R)$ algebra \cite{KM}.
Therefore one might argue that an exact solvability for other energy
values would still be possible. Nevertheless, the possibility of further 
polynomial solutions can be excluded by recent result by Zhang \cite{Zh}.
Indeed, let us consider the differential equation
\begin{equation} 
\left[ X(z)\frac{d^2}{dz^2}+Y(z) \frac{d}{dz}+Z(z)\right] \Psi(z)=0, 
\label{zheq}
\end{equation}
where $X(z)=\sum s_k z^k$, $Y(z)=\sum t_k z^k$, 
$Z(z)=\sum v_k z^k$ are polynomials of 
degree at most $4$, $3$, $2$, respectively.
Zhang \cite{Zh} found all polynomials $Z(z)$ such that 
Eq. (\ref{zheq}) has polynomial solutions 
$\Psi(z) = \prod_{j=1}^l (z - z_j)$ of degree 
$l$ with distinct roots $z_j$.
Theorem 1.1 of Zhang \cite{Zh} yields an algebraic 
conditions on each of the expansion
coefficients $v_k$, $k=0,1,2$ of $Z(z)$ in terms of
the expansion coefficients of $X(z)$, $Y(z)$, and of the roots $z_j$.
In the case of the Schweber's equation 
for the Rabi model [Eq. (3.23) of Ref. \cite{Schw}), 
and upon taking into account 
that Schweber's $\kappa$ is twice of ours,
\begin{eqnarray}
\lefteqn{
z(z-2\kappa)\frac{d^2 \Psi(z)}{dz^2} 
}
\hspace*{7cm}
\nonumber\\
+\left[2(\kappa\zeta-\kappa)+
(1-2\zeta+4\kappa^2) z-2\kappa z^2
\right]\frac{d \Psi(z)}{dz}
&& \hspace*{1.5cm}
\nonumber\\
+\left[\zeta^2 - \Delta^2 - 2\kappa\zeta(1-z)  \right] \Psi(z) =0.
&& \hspace*{0.4cm}
\label{rmde}
\end{eqnarray}
Because for the Rabi model Zhang's $s_4=s_3=t_3=0$,
Zhang's condition (1.8) on $v_2$ reduces to $v_2=0$, 
and Zhang's condition (1.9) on $v_1$ reduces to $v_1=-lt_2$, or,
$\zeta=l$. The latter leads to Eq. (\ref{bslnc}), i.e., 
again to the Juddian exact isolated solutions \cite{Jd,Ks,KL}. 
One arrives at the same conclusion if Zhang's conditions \cite{Zh}
are applied to Eq. (5) of \cite{KKT}, which is another
variant of (\ref{rmde}).
Therefore, the Rabi model has no other polynomial solution than 
the Juddian exact isolated solutions.
Any exact nondegenerate solution of the Rabi model is 
characterized by infinite set of nonzero expansion coefficients
$\phi_n$, which for sufficiently large $n$ behave as
$\phi_n \sim (-\kappa)^n/n!$ [cf. the Perron-Kreuser theorem (\ref{mins}) and 
the recurrence Eq. (\ref{rbmb2})].
It is not possible to have $\phi_n\equiv 0$ for $n>l$, where $l$
is some positive constant.
Indeed, Eq. (\ref{rbmb2}) reduces for $n=l+1$ to
$\phi_l/(l+2) =0$, in contradiction to that $\phi_l\ne 0$.
Only if two such solutions become degenerate,
a linear combination of the solutions could result in a Juddian exact isolated
analytic solution characterized in that $\phi_n\equiv 0$ for $n>l$.

\subsection{Zeros of $\phi_n$ and the spectrum}
\label{sc:phn}
In the case of the displaced harmonic oscillator,
the solution of the recurrence for $\phi_n$ 
was given by Eq. (\ref{cnsp}). In general, rapid growths of 
the product $\prod_{k=0}^{n} (\zeta-k)$ in (\ref{cnsp}) 
essentially cancels out the $1/n!$ prefactor, leads
to a {\em finite} radius of convergence of the series 
for $\upvarphi(z)$ in Eq. (\ref{pss}), and prevents 
$\upvarphi(z)$ from being an element
of $\mathfrak{b}$. The points of the spectrum 
are characterized by a sudden collapse of the
degree of $\phi_n$, which is in general
polynomial of degree $n$ in energy, 
to a polynomial of merely the $(l-1)$th order for any $n\ge l$
at the $l$th spectral point (including $l=0$).
Indeed, at the points of the spectrum 
$\zeta=\upepsilon+\kappa^2=j\in \mathbb{N}$ 
the sum over $j$ in Eq. (\ref{cnsp}) runs only between
$j=0$ and $j=l-1$ for $n\ge l$. Otherwise 
the product on the right-hand side
of Eq. (\ref{cnsp}) vanishes.  
The leading $(l-1)$th order in $\zeta$ 
is rapidly decreasing with increasing $n$ as 
\begin{equation}
(-1)^{n+1-l} \frac{l \kappa^{n+2-2l}}{(n+1-l)!},
\end{equation}
which implies $\upvarphi(z)\in\mathfrak{b}$ \cite{Schw}. 
Thus the very same product terms
$\prod_{k=0}^j(\zeta-k)$, which initially
prevented $\upvarphi(z)$ from being an element
of $\mathfrak{b}$, come later on to the rescue 
and ensure that $\upvarphi(z)\in\mathfrak{b}$ 
for the spectral points 
$\zeta\in\mathbb{N}$ (including $\zeta=0$).

It appears plausible that $\phi_n$'s can be expressed as
a sum of such product terms involving the spectrum 
also in general case (e.g. for the Rabi model), 
and a point of the spectrum would then manifests 
itself by a sudden collapse of the
degree, and magnitude, of $\phi_n$. 
The latter is necessary, because 
$\phi_n$ has to vanish in the limit 
$n\rightarrow\infty$ at the points
of the spectrum in order to guarantee that 
$\upvarphi(z)\in\mathfrak{b}$. In the case of 
the displaced harmonic oscillator, the $\phi_n$'s 
are generated by the recurrence (\ref{rbmb2h})
as polynomials in $x$. The leading order of
$\phi_n$ in $x$ is then provided by the polynomial term
\begin{equation}
\frac{1}{n!}\, \prod_{l=0}^{n-1} \left(x-\frac{l}{\kappa}\right).
\end{equation}
However, starting from the recurrence (\ref{rbmb2}), it is 
highly nontrivial to arrive at Eq. (\ref{cnsp}), and
hence to identifying the spectrum, even in
the exactly solvable case. Such a step from (\ref{rbmb2}) 
to (\ref{cnsp}) is established neither 
here nor by Braak \cite{Br}.

\subsection{Compatibility}
\label{sc:cmb}
With increasing $n$, the PFD  (\ref{pfdrl}) 
defines a sequence of
rational functions with simple real poles and positive residues.
Kritikos ($\S4$ of Ref. \cite{Krt}) showed that if the sequence 
of such rational functions
\begin{equation}
R_n(z)=\sum_{k=1}^n \frac{M_{nk}}{z-x_{nk}}
\end{equation}
converges {\em uniformly} in a proximity of some point $z_0\in\mathbb{C}$, 
then the sequence converges everywhere in the complex
plane with a possible exception of the real axis.
The convergence is {\em uniform} in any bounded region
of the complex plane with a nonzero distance from the 
real axis. The latter is obviously compatible with our main result.

In general one cannot always guarantee that, 
such as in Eq. (\ref{rsfrm}), 
$M_{nk}\rightarrow {\cal M}_k=d\psi>0$ in the 
limit $n\rightarrow\infty$. 
Indeed, Theorem I in $\S8$  of Grommer \cite{Gr}
merely ensures that for any $m>0$
\begin{eqnarray}
\psi_n(x_{nk}-0) &<&  \psi_{n+m} (x_{nk})  < \psi_n(x_{nk})
\nonumber\\
  &=& \psi_n(x_{n,k+1}-0) <  \psi_{n+m} (x_{n,k+1}),
\hspace*{1cm}
\label{grs1}
\end{eqnarray}
where as usual $\psi_n(x-0)$ denotes the 
left-side limit of $\psi_n$ at $x$. 
On taking the limit $n\rightarrow\infty$ one cannot 
exclude that $\psi(\xi_{k})=\psi(\xi_{k+1})$ for some $k$,
and hence $d\psi(\xi_{k+1})\equiv 0$.
The above Grommer's theorem appears to be related 
to the fact that, given the sharp inequality (\ref{zing}),  
$\xi_k^{(1)}$ could coincide with one of $\xi_k<\xi_{k+1}$.
Because $\xi_k^{(1)}<\xi_{k+1}^{(1)}$, one can only 
exclude that two subsequent $\xi_k^{(1)}$ 
and $\xi_{k+1}^{(1)}$
coincide with a single $\xi_k$ or $\xi_{k+1}$. 
If $\xi_k^{(1)}=\xi_{k+1}$,
then $d\psi(\xi_{k+1})=0$ in the 
{\em Mittag-Leffler} PFD (\ref{rpthml}).
However, the latter can be prevented in the case of the
Rabi model [cf. Eq. (\ref{3p4p1l})].

\subsection{Open problems}
\label{sc:open}
The dynamics and long-time evolution of the Rabi model 
is well understood only for rather weak 
couplings $\kappa=g/\omega\lesssim 10^{-2}$, where the Rabi model 
can be reliably approximated by the JC model \cite{JC,SK}. 
The latter was originally proposed as 
an exactly-solvable approximation to
the Rabi model by applying the RWA and neglecting
rapidly oscillating counterrotating terms \cite{JC,SK}.
The JC model provides the basis for, 
from the Rabi model perspective 
somewhat misleadingly called, {\em strong-coupling} regime 
\cite{CRL,KGK,CC}. The latter encompasses the
cavity quantum electrodynamics and associated with it 
vacuum-field Rabi oscillations 
of atoms, molecules, and quantum-dots in a cavity \cite{KGK}. 
Long-time behavior of various dynamical variables 
of the JC model can be described by analytic approximations \cite{ENS}
and the dynamics shows periodic spontaneous collapse and revival 
of coherence \cite{ENS,Agr}.
With new experiments rapidly approaching the limit of
the deep strong coupling regime
characterized by $\kappa\gtrsim 1$ \cite{CRL}, 
the question of major physical relevance is that of
the dynamics of the Rabi model for $\kappa\gtrsim 0.1$ 
\cite{CRL,FL,WoK,WVR}. 
A great deal of insight into the dynamics of the Rabi model 
has been gained by Casanova et al. \cite{CRL} 
in the limit $\omega_0/\omega\ll 1$ by means of 
an expansion in the small parameter $\omega_0/\omega$. 
Photon wave number packets were shown to propagate coherently 
along two independent parity chains of states 
and, like in the JC model \cite{ENS,Agr}, exhibited
a collapse-revival pattern of the system population \cite{CRL}.
Nevertheless, still only very little is known
about the dynamics in a very interesting 
physical region of $\omega_0/\omega\simeq 1$ \cite{CRL,FL,WoK,WVR}.
The just established link between the Rabi model and 
the OPS's could improve calculations of the eigenvalues
and eigenfunctions, which are prerequisite for
a reliable calculation of the dynamics of the Rabi model,
and shed light on its long-time evolution 
for all values of the dimensionless coupling $\kappa$ \cite{CRL}.
Indeed, it can be shown that the number of calculated energy levels 
is almost {\em two orders} of magnitude higher \cite{AMtb} 
than is possible to obtain by means of the alleged 
``analytical solution" 
of the Rabi model \cite{Br}.
The only numerical limitation in calculating zeros 
are {\em over-} and {\em underflows} in double precision. 
Typically, with increasing $n$ the respective 
recurrences yield first increasing and then decreasing 
$\phi_n$, $A_n$, and $B_n$. By using an elementary 
stepping algorithm this limits the total number
of energy levels that can be 
determined to ca. {\em 1350} levels
per parity subspace, or to ca. {\em 2700} levels for the 
Rabi model in total in double precision \cite{AMtb}.
It is conceivable that on using more sophisticated algorithm
the number of calculable levels could be comparable
to that obtained by numerical schemes involving 
diagonalization of tridiagonal matrices.

The nearest-neighbor energy level spacing distribution 
is customarily used to distinguish between integrable 
models and chaotic systems
\cite{BT,BGS,ZR}. Kus \cite{Ksc} observed 
that neither Poissonian nor Wigner distributions can 
describe the level statistics for the Rabi model. 
Instead Kus \cite{Ksc} found a nongeneric distribution 
of a ``picket-fence" type.
Our recent results suggest that the level statistics 
should resemble that of the zeros 
of suitable orthogonal polynomials \cite{AMtb}. 
At the same time, the significantly larger number
of computable energy levels within our approach enables 
one to perform a refined statistical analysis of the spectrum.

Carleman's criterion states that the Hamburger 
moment problem associated with the positive-definite OPS 
(\ref{chi3tr}) and (\ref{chi3tric}) is {\em determined}
(i.e., the Stieltjes weight function $\psi$ is unique),
if the condition (\ref{crlcr}) is satisfied \cite{Chi}.
An alternative sufficient condition 
for the determinacy of the classical Hamburger moment problem
is that the moments (\ref{mf}) of the positive moment functional 
${\cal L}$ have to satisfy 
(cf. Eq. (3) of Ref. \cite{Hb}; $\S5$ of Ref. \cite{Hb0})
\begin{equation}
|\mu_n|\le \frac{\theta}{\Lambda^n}\, n!,
\label{hmbc}
\end{equation}
where $\theta$ and $\Lambda$ are two real positive constants.
Hamburger's condition (\ref{hmbc}) is also sufficient
condition for the {\em uniform} convergence of the
associated infinite continued fraction in any {\em closed} domain
$\Omega$ of the complex plane which does not contain any part of 
the real axis \cite{Hb}.
The sequence of convergents in Eqs. (\ref{pfdr}) and (\ref{pfdrl}) then
converges {\em uniformly} to the infinite 
continued fraction in Eq. (\ref{fdfs}), 
which ensures the convergence of 
the latter \cite{Hb} (see also pp. 214-215 of Ref. \cite{Hb0}). 
This raises the question regarding a mutual relation 
between the respective asymptotic behaviors of the $\lambda_n$'s and
$\mu_n$'s. In the special case of $c_n\equiv 0$ in Eq. (\ref{chi3tr}) 
for all $n\in\mathbb{N}$, Bender and Milton \cite{BM} showed 
that $\lambda_n\sim n$ implies $\mu_n\sim n!$ and vice versa.
It would be interesting to prove rigorously if such a relation 
holds also in more general case of $c_n\ne 0$, such as for the Rabi model.

Last but not the least, an interesting question is 
if the ideas presented here could also be extended
to systems characterized by a 
three-term difference equation 
(\ref{3trg}) with {\em periodic} coefficients, like in the 
Hofstadter problem of Bloch 
electrons in rational magnetic fields \cite{Hf}.

\section{Conclusions}
\label{sec:conc}
On applying the theory of orthogonal polynomials, 
the eigenvalue equation and eigenfunctions of 
the quasi-exactly solvable Rabi model
was shown to be determined in terms of three systems of monic 
orthogonal polynomials. The formal Schweber 
quantization criterion for an energy variable $x$, 
originally expressed in terms of infinite continued
fractions, was shown to be equivalent to
a meromorphic function $F(x)$ in the complex 
plane $\mathbb{C}$. $F(x)$ can be expressed by the partial fraction 
decomposition (\ref{rpthml}) with {\em real simple} poles and 
positive residues ${\cal M}_k$ defined by (\ref{rsfrm}).
Thereby the calculation of spectrum corresponding 
to the zeros of $F(x)$ was greatly facilitated. 
One obtains at once that (i) $F(x)$ monotonically 
decreases from $+\infty$ to $-\infty$ 
between any two of its subsequent poles $\xi_k$ and $\xi_{k+1}$,
(ii) there is exactly one zero of $F(x)$ for $x\in (\xi_k,\xi_{k+1})$, 
and (iii) the spectrum corresponding to the zeros of $F(x)$ 
does not have any accumulation point. Additionally, one can 
provide much simpler proof of that the spectrum in each parity eigenspace 
${\cal B}_\pm$ is necessarily {\em nondegenerate}.
Recent claims regarding solvability and integrability of 
the Rabi model \cite{Br,Sln} were critically examined.
Compatibility of our results with 
some other results of the theory of infinite 
continued fractions and complex analysis was explicitly demonstrated.

\section{Acknowledgment}
I like to thank anonymous referees for valuable comments.
Continuous support of MAKM is largely acknowledged.

\appendix

\section{Technical remarks}
\label{app:tchr}
Assuming that the power series
$\sum_{n=0}^\infty \mu_n/z^n$ has a nonzero radius 
of convergence $|z|>R>0$, i.e., much weaker condition than
Hamburger's condition (\ref{hmbc}), 
Grommer arrived from an integral representation
(cf. Eq. (63) of Ref. \cite{Gr})
\begin{equation}
f(z)=\int_{-R}^R\frac{d\psi(x)}{z-x}
\label{grint}
\end{equation}
to the {\em Mittag-Leffler} PFD (cf. Eq. (66) of Ref. \cite{Gr})
\begin{equation}
f(z)=\sum_{j=1}^\infty \frac{{\cal M}_j}{z-\xi_j}\cdot
\label{mlpfd}
\end{equation}
Essential to his arguments was that $f(z)$ 
remained finite and different from zero
for any nonreal $z\in\mathbb{C}$. However, in virtue of
\begin{equation}
\int_{-\infty}^\infty \frac{dx}{(u-x)^2+q^2} =  \frac{\pi}{q} < \infty,
\end{equation}
the very same is also true if the integration range
in Eq. (\ref{grint}) were, such as in our case, extended to infinity.

By making use of the Nevanlinna theorem \cite{N}
(later rediscovered by Sokal \cite{So} in his 
improvement of Watson's theorem on Borel summability;
for an extension of the Nevanlinna-Sokal theorem to differently 
shaped region see Ref. \cite{AMcmp}), 
which was not known to Hamburger at the time he wrote his \cite{Hb}, 
one can immediately amend Hamburger theorem under his item 4
on pp. 33-34 of \cite{Hb}. One can prove that Hamburger's function $f(z)$
is asymptotically represented by the power series $S(z)$ 
not merely in an angular domain $\varepsilon\le \arg z\le \pi-\varepsilon$,
where $\varepsilon$ an arbitrary small positive infinitesimal, but rather
in a disc tangent to the real axis. 

The zeros of $P_n(x)$ can be associated with eigenvalues of 
{\em Jacobi} matrices. The latter are derived from the 
coefficients of the recurrence (\ref{chi3tr}) 
upon writing each $\lambda_n>0$ as a product 
of two complex conjugate numbers
$\lambda_n=z_nz_n^*$, and assembling them into the Hermitian matrix
\begin{equation}
M_n=\left[
\begin{array}{ccccccc}
c_1   & z_2 & 0 & \ldots  & 0 & 0 & 0
\\
z_2^* & c_2 & z_3 & \ldots & 0 & 0 & 0
\\
0 & z_3^*   & c_3 & \ldots & 0 & 0 & 0
\\
\cdot  & \cdot & \cdot & \ldots & \cdot & \cdot & \cdot
\\
0    & 0  & 0 & \ldots & z_{n-1}^* & c_{n-1} & z_n
\\
0    & 0  & 0 & \ldots & 0 & z_{n}^* & c_n
\end{array}
\right].
\end{equation}
Then the eigenvalues of $M_n$ are the zeros of $P_n(x)$.
(Ex. I-5.7 on p. 30 of Ref. \cite{Chi}).

Using the results of Hamburger \cite{Hb}, one can prove that
\begin{equation}
w_n(t)= \sum_{k=1}^n M_{nk}\, e^{x_{nk} t}
\label{vnt}
\end{equation}
converges uniformly toward 
\begin{equation}
W(t)= \sum_{n=0}^\infty \frac{\mu_n}{n!}\, t^n
\label{vt}
\end{equation}
within any vertical stripe $|t|\le R-\delta$,
where $R>0$ is a nonzero radius of convergence 
of the series for $W(t)$ and 
$\delta$ is some infinitesimally small number.
The continued fraction in Eqs. (\ref{fdfs}) and
(\ref{r0l}) converges then uniformly to 
the Borel transform 
\begin{equation}
F(z)= a_0 + \int_0^{-i\infty} W(t)\, e^{-tz}\, dt
\end{equation}
in any closed domain of the complex plane with Im $z\ge \delta>0$ 
(cf. $\S3$ of Ref. \cite{Hb}).

\vspace*{2.5cm}


\end{document}